\newif\ifsubmit
\submittrue

\documentclass[twocolumn,showpacs,preprintnumbers,amsmath,amssymb,nofootinbib,superscriptaddress]{revtex4}

\ifsubmit
\usepackage{graphicx}
\else
\usepackage{graphicx}
\fi

\newcommand {\MAYA} {\textsc{Maya}}

\begin{document}

\title{Black hole head-on collisions and  gravitational waves with fixed mesh-refinement
and dynamic singularity excision}

\author{Ulrich Sperhake}
\affiliation{Center for Gravitational Wave Physics, \\
             Penn State University, University Park, PA 16802}

\author{Bernard Kelly}
\affiliation{Center for Gravitational Wave Physics, \\
             Penn State University, University Park, PA 16802}
\affiliation{Center for Gravitational Wave Astronomy, \\
             University of Texas at Brownsville, Brownsville, TX 78520}

\author{Pablo Laguna}
\affiliation{Center for Gravitational Wave Physics, \\
             Penn State University, University Park, PA 16802}
\affiliation{Institute for Gravitational Physics \& Geometry \\
             Departments of Astronomy \& Astrophysics and Physics \\
             Penn State University, University Park, PA 16802}

\author{Kenneth L. Smith}
\affiliation{Center for Gravitational Wave Physics, \\
             Penn State University, University Park, PA 16802}

\author{Erik Schnetter}
\affiliation{Max-Planck-Institut f\"ur Gravitationsphysik,
             Albert-Einstein-Institut,
             14476 Golm, Germany}

\date{\today}

\begin{abstract}
  We present long-term-stable and convergent evolutions of head-on
  black hole collisions and extraction of gravitational waves
  generated during the merger and subsequent ring-down.
  The new ingredients in this work
  are the use of fixed mesh-refinement and
  dynamical singularity excision techniques.
  We are able to carry out head-on collisions
  with large initial separations and demonstrate that
  our excision infrastructure is capable of accommodating the motion of
  the individual black holes across the computational domain as well as
  their merger.
  We extract gravitational waves from these simulations using the
  Zerilli-Moncrief formalism and find the ring-down radiation
  to be, as expected,
  dominated by the $\ell=2$, $m=0$ quasi-normal mode. The total
  radiated energy is about $0.1\,\%$ of the total ADM mass of the system.
\end{abstract}

\pacs{04.25Dm, 04.30Db}

\maketitle

\section{Introduction}
\label{sec: introduction}

One of the most urgent problems currently under investigation in
general relativity is the coalescence of binary compact objects. The
importance of this scenario largely arises from its significance for
the new research field of gravitational wave astronomy; ground-based
gravitational wave interferometric detectors, such as LIGO, GEO600,
VIRGO and TAMA, have started collecting data.
The in-spiral and merger of neutron stars and black holes
is considered among the most promising sources of gravitational radiation
to be detected by these instruments.
Theoretical predictions of the resulting wave patterns will be
important both for the detection and for the physical interpretation
of such measurements. Although approximation techniques have been
remarkably successful, the theoretical calculation of wave forms
from compact object binaries will most likely require the numerical solution of
the Einstein field equations.

Very important milestones have been achieved recently in
the numerical modeling of compact object binaries.
For instance, a few orbits of binary neutron stars are now possible
(see e.g.\,\cite{Shibata2003,Marronetti2004,Miller2004}). Also, with binary black holes,
Br\"ugmann et al.\,\cite{Bruegmann2004}
have been able to carry out evolutions lasting for timescales longer than
one orbital period.
These varied efforts in modeling compact binary systems with numerical
relativity share a common goal -- to extract accurate waveform templates
(see \cite{Miller2005} for a study on accuracy requirements
of such waveform calculations).
This ``Holy Grail'' of numerical relativity, however,
still eludes the community.
A unique problem faced in the simulation of black hole spacetimes
is the presence of physical singularities, which makes this task
particularly delicate from a numerical point of view.

For a long time after the pioneering work of Eppley and Smarr in the 1970s
\cite{Eppley1975, Smarr1976, Smarr1979},
progress in numerical black hole evolutions was rather slow, and
the codes suffered from instabilities after relatively short times.
In contrast, the last few years have seen a change of strategy in
the area of numerical relativity. More emphasis has been put on
analyzing the structure of the evolution equations underlying the
numerical codes (see e.g.\,\cite{Shibata1995, Friedrich1996,
Frittelli1996, Anderson1999, Baumgarte1999, Alcubierre2000b,
Kidder2001, Yoneda2002, Sarbach2002b, Bona2003a}).
In combination with increased computational
resources and advanced treatments of the spacetime singularities, this
has led to vastly improved stability properties in the evolution of
spacetimes with a single black hole either at rest or moving
through the computational domain
\cite{Alcubierre2001b,Yo2002,Alcubierre2003b,Sperhake2004}.
In view of these encouraging results, attention has been shifting again toward
the application of these techniques to the evolution and merger of
binary black holes (see
e.g.\,\cite{Brandt2000, Alcubierre2003b, Bruegmann2004, Alcubierre2004, Alcubierre2004b}).

In this paper, we will focus on the numerical evolution of head-on
collisions of black holes implemented in the framework of dynamical singularity
excision. The study of head-on colliding black holes has attracted a
good deal of attention in the past, both numerically and analytically.
We have already mentioned the early work of Eppley and Smarr who
numerically evolved Misner data for equal-mass black hole collisions
in axisymmetry. The waveforms extracted from these evolutions differ
little in amplitude and phase from those obtained in perturbative
calculations \cite{Smarr1979}. The head-on collision was
re-investigated numerically in the 1990s
\cite{Anninos1993, Anninos1995c, Anninos1995d, Baker1997, Anninos1998} with
generally good agreement between numerical results and those obtained
from perturbation theory for various initial configurations. As a
whole, the approximate analytic study of binary black hole head-on
collisions
has been remarkably successful and has generated a wealth of
literature. The late stage of the collision, when the separation of
the two holes is sufficiently small, is well described by the {\em
close limit} approximation (see e.g.\,\cite{Price1994, Abrahams1994}
as well as \cite{Gleiser1996} for an estimate of the quality of the
approximation using second-order perturbation theory). Similarly the
particle approximation has been used to describe the in-fall phase with
large black-hole separation either in the framework of post-Newtonian
theory (see e.g.\,\cite{Simone1995}) or using Newtonian trajectories
with appropriate modifications \cite{Anninos1995c, Anninos1995d}. In
particular, Anninos et al.\,\cite{Anninos1995d}
combined their ``particle-membrane''
technique with results from close limit calculations to provide
estimates of the total radiated energy that are in good agreement with
the results of numerical relativity over the whole range of initial
black hole separations (see their Fig.\,11).

The encouraging results
of these perturbative approximations gave rise to the idea of
combining them with full numerical evolutions via matching on slices
of constant time. With this temporal matching, it was possible to
extend the life of simulations of binary
black holes which would otherwise
stop because of numerical instabilities.
This is now commonly known as the ``Lazarus'' approach;
developed by Baker et al.\,\cite{Baker2000} it has demonstrated
good waveform agreement
with non-linear evolutions.

The general picture that has
emerged in analytic as well as numerical studies of equal mass
head-on collisions is that gravitational
wave emission is dominated by the $\ell=2$, $m=0$ quasi-normal
oscillations of the post-merger black hole and that the total
radiated wave energy is typically
less than a percent of the system's ADM mass.

In spite of their limited astrophysical relevance,
simulations of black hole
head-on collisions still represent an important problem.
In view of the eventual target, the in-spiral and merger of
a binary black hole, these simulations constitute
a key milestone in the development of numerical codes,
a natural next
step after successful evolutions of single black holes.
While numerical codes have continued to
improve in the simulation of head-on or grazing collisions
\cite{Bruegmann1999, Brandt2000, Alcubierre2001c}, long-term-stable
simulations have only been obtained rather recently by
Alcubierre et al.\,\cite{Alcubierre2003b} using
the puncture method. In
\cite{Alcubierre2004} the combination of this approach with
the ``Simple Excision'' technique of Alcubierre and Br\"ugmann
\cite{Alcubierre2001b} has
been shown to preserve the long-term stability and yield good agreement
in the wave forms. Almost parallel to this work black hole
head-on collisions have been simulated in the framework
of mesh refinement \cite{Fiske2005} and fourth-order
finite differencing techniques \cite{Zlochower2005}.
In this paper we
present the first such long-term-stable evolutions obtained
in fixed mesh-refinement as well as using
the dynamical singularity excision technique.

The simulations presented in this paper have been obtained with the
{\sc Maya} evolution code. This code and in particular
its dynamical singularity excision
technique are described in detail in \cite{Shoemaker2003} (referred to
as Paper 1 from now on). A key development in recent months has been the
combination of the {\sc Maya} code with the mesh-refinement package
\textsc{Carpet} \cite{Schnetter2004}. \textsc{Carpet} has enabled us to push the
outer boundaries substantially further out than was possible with
the uni-grid version of {\sc Maya}; this was crucial in the extraction
of gravitational wave signals.

The paper is organized as follows:
The construction of binary black hole initial data used
in this work is described in Sec.\,\ref{sec: initial data},
while the details of the numerical evolution scheme
are given in Sec.\,\ref{sec: numerical evolution}. The gravitational
waveforms are analyzed in Sec.\,\ref{sec: results} with regard to their
convergence properties and the physical information they provide.
We conclude in Sec.\,\ref{sec: conclusion} with a discussion of our
results and the implications for future work. In the appendix we
present the details of the wave extraction using the Zerilli-Moncrief
formalism on a single black hole background in Kerr-Schild coordinates.
Throughout the paper, units are given in
terms of $M$, with $M = \,^Am+\,^Bm$, the sum of the bare masses of the
individual black holes in the binary.


\section{Initial data}
\label{sec: initial data}

The construction of astrophysically meaningful initial data for
simulations of compact binaries is a highly non-trivial problem
and represents an entire branch of research in numerical relativity
(see \cite{Cook2000} for a detailed overview).

For the simulations discussed in this work, we construct initial
data according to the procedure of Matzner et al.\,\cite{Matzner1998},
referred to from now on as ``HuMaSh'' data.
This type of data has been used, for example, in
binary grazing-collision evolutions by Brandt et al.\,\cite{Brandt2000}.
The starting point of the HuMaSh data is
the Kerr-Schild spacetime metric
\begin{equation}
    g_{\mu\nu} = \eta_{\mu\nu} + 2\,H\,l_\mu\,l_\nu\,,
\end{equation}
where $\eta_{\mu\nu}$ is the Minkowski metric, and $l_\mu$ is
a null-vector with respect to both $g_{\mu\nu}$ and $\eta_{\mu\nu}$.
In terms of 3+1 quantities, the Kerr-Schild metric is given by
\begin{eqnarray}
\label{ID:KS_g}
    \gamma_{ij} &=& \delta_{ij} + 2\,H\,l_i\,l_j\,,\\
\label{ID:KS_lapse}
    \alpha &=& (1+2\,H\,l^t\,l^t)^{-1/2}\,,\\
\label{ID:KS_shift}
    \beta^i &=& - \frac{2\,H\,l^t\,l^i}{1+2\,H\,l^t\,l^t}\,.
\end{eqnarray}
In particular, for a non-rotating black hole of mass $m$ at rest at the origin
\begin{eqnarray}
    H &=& \frac{m}{r}\,,\\
    l_{\mu} &=& (1,\vec{l})\,,\\
    \vec{l} &=& \frac{\vec{x}}{r}\,,
\end{eqnarray}
with $r = \sqrt{\delta_{ij}\,x^i\,x^j}$.

As pointed out in Ref.~\cite{Matzner1998}, Kerr-Schild coordinates
have two attractive features.
First, with relevance to singularity excision, the coordinates penetrate
the horizon.
Second, the Kerr-Schild spacetime metric is
form-invariant under a boost transformation.
Specifically, under a boost $v$ along the $z$-axis, all
that changes is
\begin{eqnarray}
    H &=& \frac{m}{R}, \nonumber\\
          l_{\mu}&=& \left[ \frac{\gamma\,(R - v\,Z)}{R},
          \frac{x}{R} , \frac{y}{R} ,
          \frac{\gamma\,(Z - v\,R)}{R} \right], \nonumber
\end{eqnarray}
where $\gamma=1/\sqrt{1-v^2}$, $Z \equiv \gamma (z - v \, t)$,
and $R^2 \equiv x^2 + y^2 + Z^2$.

A spatial metric approximating initial data of
two black holes, with bare masses  $\,^Am$ and $\,^Bm$
respectively, can be given by
\begin{eqnarray}
\label{ID:HuMaSh_g}
  \gamma_{i j} & = & \,^A \gamma_{i j} + \,^B \gamma_{i j} - \delta_{i j}\,,
\end{eqnarray}
where the $\,^A \gamma_{i j}$ and $\,^B \gamma_{i j}$ are to be understood as
the Kerr-Schild spatial metric (\ref{ID:KS_g})
evaluated at the corresponding black hole coordinate
position and boost velocity.
Following the HuMaSh prescription, the extrinsic curvature is constructed
from
\begin{eqnarray}
\label{ID:HuMaSh_K}
K^i\,_j & = &  \,^A K^i\,_j +  \,^B K^i\,_j\,,
\end{eqnarray}
where the $\,^AK^i\,_j $ and $\,^BK^i\,_j $ are also to be understood as
the Kerr-Schild extrinsic curvature
evaluated at the corresponding black hole coordinate position
and boost velocity.

It is important to point out that the HuMaSh data do not provide a close limit;
that is, as the coordinate separation of the black holes shrinks to zero,
the resulting metric is not that of a single black hole with
mass $M = {}^Am + {}^Bm$.
Furthermore, as given by (\ref{ID:HuMaSh_g}) and (\ref{ID:HuMaSh_K}),
the HuMaSh data do not
satisfy the Einstein constraints.
As suggested by Marronetti et al.
\cite{Marronetti2000b},
one could attenuate the constraint violations of the
HuMaSh data using an exponential damping factor.
Bonning et al. \cite{Bonning2003} used this attenuating
procedure and provided a fairly
detailed investigation of the general properties of the underlying
HuMaSh data. In particular, they show that it has the correct
gravitational binding energy in the Newtonian far-separation limit.

By construction the HuMaSh data will satisfy the constraint equations
only in the limit of infinite separation of the two black holes. A
rigorous treatment of a scenario with finite separation would thus
require using the HuMaSh data as freely-specifiable background data
and then solving the Hamiltonian and momentum constraint
according to York's initial data prescription \cite{York1979}.
Since one of our primary goals is to extract gravitational waves from
the black hole simulations, we need a sufficiently large computational
domain, only attainable via mesh-refinement.
Unfortunately, elliptic solvers operating on numerical grids with various
refinement levels have not yet been developed in our {\sc Maya} code.
Using the uni-grid solvers available, on the other hand,
is computationally
prohibitive. Therefore, as with the grazing collision work
in Ref.~\cite{Brandt2000}, the simulations in this paper were obtained
using the constraint-violating HuMaSh initial data.

In this context one has to keep in mind that the constraints will
always be violated at the level of numerical accuracy in a real
simulation.
\begin{figure}
  \begin{center}
    \includegraphics[height=270pt,angle=-90]{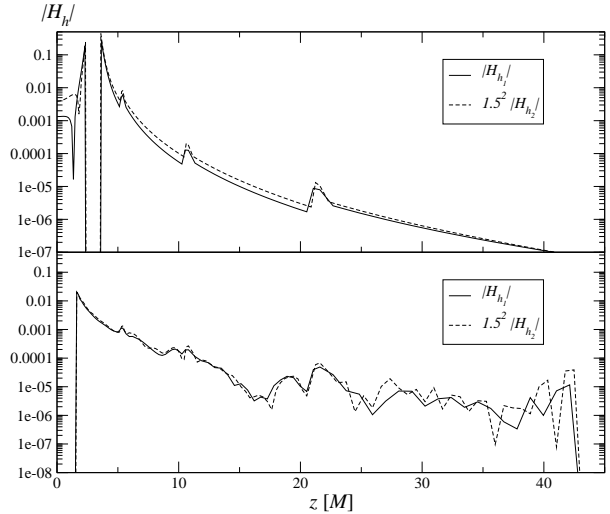}
    \caption{Convergence analysis of the Hamiltonian constraint
             along the $z$-axis at $t=0$ (upper panel) and $t=25.92\,M$
             (lower panel).}
    \label{fig: zham}
  \end{center}
\end{figure}
In consequence, the constraint violations inherent to the HuMaSh
construction are problematic only if they are significant
relative to the limit set by finite grid resolutions
in the course of the evolution. Furthermore,
the constraint violations
are reduced as we increase the initial separation
of the black holes.

In order to estimate the significance of the constraint violations
in the HuMaSh data relative to the numerical evolution errors,
we have analyzed the convergence properties of the constraint
violations in the course of the simulations. Because the
initial data are not exactly constraint satisfying, we do not expect
the values of the constraints to converge to zero at second order
in the continuum limit. Let $\mathcal{H}_0$ denote the Hamiltonian constraint
violation at the continuum level inherent to the HuMaSh data.
$\mathcal{H}_0$ is thus independent of the grid resolution.
For a second-order convergent
evolution code, the numerical constraint
violation $\mathcal{H}_h$ will be given by
\begin{equation}
  \mathcal{H}_h = \mathcal{H}_0 + c\,h^2\,,
\end{equation}
where $h$ denotes the grid resolution and $c$ is a function
of space and time, independent of $h$.
We have carried out two head-on collisions with black holes separated by a
coordinate distance $6\,M$. These runs have
resolutions $h_1 = 0.135\,M$ and
$h_2=h_1/1.5\,M$ on the finest refinement level respectively, with the
grid spacing doubled in each of the three additional coarser levels.
(These are the ``coarse'' and ``medium'' simulations in the convergence
analysis of the waveforms in Sec.\,\ref{sec: convergence}.)
For a second-order convergent code we expect the constraint violations
at the two resolutions to obey the relation
\begin{equation}
  q \equiv
  \frac{\mathcal{H}_{h_1}}{\mathcal{H}_{h_2}} = \frac{\mathcal{H}_0 + c\,h_1^2}
      {\mathcal{H}_0 + c\,h_2^2}.
\end{equation}
It is instructive to consider the following two limiting cases of this equation.
First, if the contribution $\mathcal{H}_0$ due to the use of initially
unsolved data dominates the total
constraint violation, we obtain $q = 1$ and should observe no convergence
at all. On the other hand, if $\mathcal{H}_0$ is negligible,
we expect second-order convergence and $q=1.5^2$.
\begin{figure}
  \begin{center}
    \includegraphics[height=270pt,angle=-90]{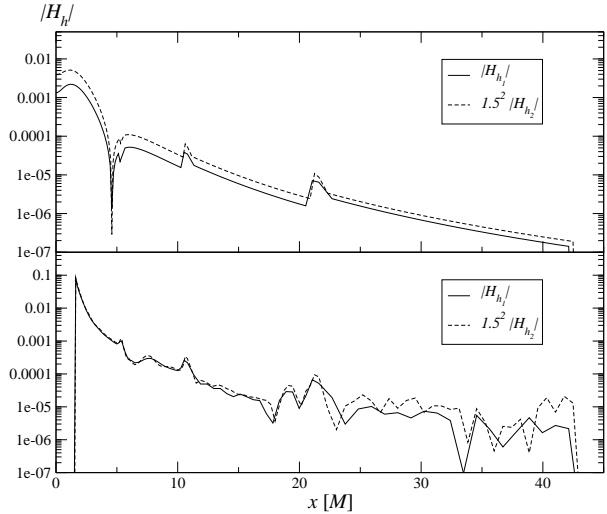}
    \caption{Convergence analysis of the Hamiltonian constraint
             along the $x$-axis at $t=0$ (upper panel) and $t=25.92\,M$
             (lower panel).}
    \label{fig: xham}
  \end{center}
\end{figure}

In view of this result, we have investigated the point-wise convergence
of both the Hamiltonian and the momentum constraints.
Fig.\,\ref{fig: zham} shows
the Hamiltonian constraint violation $\mathcal{H}_h$ along
the $z$-axis (collision axis) for
both runs, with the medium resolution result
($h_2=h_1/1.5$) scaled by a factor
of $1.5^2$.
The upper panel shows $\mathcal{H}_h$ at
$t=0$ and the lower panel at $t=25.92\,M$.
It is clear from this figure that at $t=0$
second-order convergence is obtained in the strong field
area close to the hole as well as
in the far-field limit. At intermediate distances,
on the other hand, the convergence deteriorates,
in particular near $z=0$, the region between the two holes.
In the course of the evolution, however, the convergence approaches
second order everywhere except in the domain of influence of
the outer boundary, as clearly
demonstrated in the lower panel of Fig.\,\ref{fig: zham}.
The constraint violations are second-order convergent out
to about $z=22\,M$.
We emphasize that these outer boundary effects
are not related to the question of initial data and have also been
observed, for example, in the puncture evolutions of
Alcubierre et al.\,\cite{Alcubierre2004}.
This viewpoint is further strengthened by the convergence analysis
of the waveforms in Sec.\,\ref{sec: convergence} which also reveal
the propagation of noise from the outer boundaries as the limiting
factor.

In order to shed more light on the constraint violations at small $z$,
we show in Fig.\,\ref{fig: xham} the Hamiltonian constraint violations
along the $x$-axis at the same times. Due to the axisymmetry in the problem,
the results along the $y$-axis are identical. Clearly the initial data do
not exhibit second-order convergence, in agreement with the
results of Fig.\,\ref{fig: zham} near $z=0$. In the course of the
evolution, however, as the discretization error increases and
dominates the total violations we do find
second-order convergence of the Hamiltonian constraint except for the
outer boundary effects mentioned above.

The results obtained for the momentum constraints show the same behavior,
though they exhibit a stronger level of noise emanating from the
refinement boundaries. We currently do not have an explanation for this
discrepancy between Hamiltonian and momentum constraint
but emphasize that the overall convergence
of the momentum constraint is still close to second order up to the
outer boundary effects.

\begin{figure}
  \begin{center}
    \includegraphics[height=270pt,angle=-90]{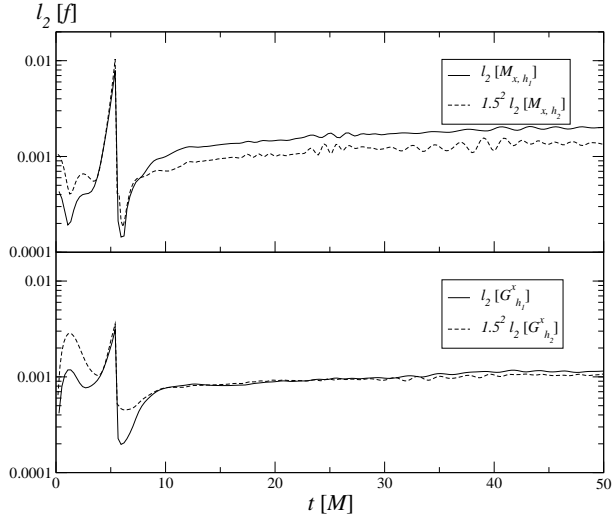}
    \caption{Convergence analysis of the $\ell_2$-norms
             as functions of time of the
             $x$-component of the momentum and auxiliary constraints.
             }
    \label{fig: l2conv}
  \end{center}
\end{figure}

Similar convergence properties are found for the $\ell_2$-norms
of the Hamiltonian constraint $\mathcal{H}$, the momentum
constraints $\mathcal{M}_i$ and the auxiliary constraint
$\mathcal{G}^i\equiv \tilde{\Gamma}^i - \tilde{\gamma}^{jk}
\tilde{\Gamma}^i_{jk}$.
In Fig.\,\ref{fig: l2conv} we show the resulting plots for
the $x$-components $\mathcal{M}_x$ and $\mathcal{G}^x$.
As before, the constraint violations obtained with fine resolution $h_2$
are scaled by the expected factor $1.5^2$.
The results for the other components of the
constraint variables are similar and overall we find convergence
close to second order after a transition time of about $10\,M$.
Outer boundary effects do not manifest themselves as clearly
in the $\ell_2$-norms as in the point-wise convergence analysis.
We attribute this feature to the fact that the norms are dominated
by the large constraint violations near the black holes.

In summary, our results confirm that the total
constraint violation is indeed dominated by the numerical
inaccuracies inevitably generated during time evolutions at currently
available grid resolutions. This analysis generalizes the studies
of Marronetti et al.\,\cite{Marronetti2000b} to include the
discretization errors arising from the time evolution; these lead
to an extension of the allowed range of grid spacings from their
limiting value of $M/4$ to about $M/8$ as used in this work.
While constraint solving of the
initial data will be desirable for future simulations with higher
resolutions, its effect on the overall constraint violations encountered
in our simulations would be marginal at most. We emphasize in this
context that this is only the case because of the absence of exponentially
growing constraint violating modes in our simulations.

\section{Numerical Evolution}
\label{sec: numerical evolution}

From the operational point of view, it is convenient
to divide a black hole head-on collision into pre-merger, merger
and post-merger stages.
In the pre-merger stage, the evolution contains two separate
apparent horizons and therefore separate excision regions.
The merger stage is signaled  by the appearance of
a single, highly distorted apparent
horizon. Initially, these deformations of the horizon
prevent the merger of
the excision masks into a single spherical excision.
The end of the merger stage is marked
by the merger of excision masks when the deformations of the
common apparent horizon have damped-out.
The main constraint on the construction of the merged mask
is that it must be completely inside the apparent horizon
and fully contain the previous two disjoint excision regions.
During the final stage (post-merger), the evolution consists of
the settling or ring-down of the resulting black hole.

There are certain
aspects besides the handling of the excision region,
such as gauge conditions, that
require a different treatment
during each of these stages. Before we discuss these differences in detail,
we address those ingredients of the numerical evolution that remain
unchanged throughout the complete simulation.

The specific implementation of the 3+1 BSSN formulation
and the black hole excision technique
used in the \MAYA\ code have been described in detail in Paper 1.
Here we provide a brief summary with particular regard to
the changes necessary for working with fixed-mesh refinement.

We evolve the Einstein equations on a set of nested non-moving,
uniform Cartesian grids with the resolution increasing by a factor
of two for each additional refinement level.
The Einstein field equations are evolved according to the BSSN 3+1 formulation
\cite{Shibata1995,Baumgarte1999}. The exact shape of the equations as
implemented in the code is given by Eqs.\,(6)-(10) in Paper 1;
in particular this
includes the additional term in the evolution of $\hat{\Gamma}^i$
introduced by \citet{Yo2002} (the free parameter $\chi$ present in
that term is set to $2/3$ for all simulations discussed in this
paper). These equations are evolved in time with the iterated
Crank-Nicholson scheme. Spatial discretization is
second-order centered finite differencing with the exception
of advection derivatives contained in the Lie derivatives;
these are approximated with one-sided
second-order stencils.
With regard to the treatment of refinement boundaries, we follow
the recipe of Schnetter et al. \cite{Schnetter2004} to avoid loss
of convergence at these boundaries. For our code, this implies the
use of a total of six buffer zones; two points (because of the
one-sided derivatives mentioned above) for each of the three
Crank-Nicholson iterations.

With the exception of gauge quantities specified algebraically as
functions of the spacetime coordinates,
we apply a Sommerfeld-like outgoing
condition at the outer boundary \cite{Shoemaker2003}.
At the excision
boundary, we extrapolate all evolution fields from the computational
interior using quadratic or, when populating previously
excised points, cubic polynomials. With regard to mesh refinement
we note that the black hole excision is actively implemented on the
finest level, only migrating to the coarser levels via the restriction
of grid functions;
consequently, we always require that the black holes remain
within the range of the finest grid.
In practice this is not a limitation
since it is precisely the black hole neighborhood where
steep gradients are present and high resolution is therefore required.

In Paper 1, the uni-grid version of this code has been shown
to produce
long-term-stable evolutions of single Schwarzschild black holes as
well as evolutions of moving black holes lasting for about $130\,M$.
In \cite{Sperhake2004} (referred to as Paper 2 from now on),
we managed to extend the life times
of these runs to at least $6000\,M$ by describing the slicing
condition in terms of a densitized lapse function [cf. Eqs.\,(6)-(8)
therein] and demonstrated second-order convergence of the code.

The remaining aspects of the numerical simulation are specific to the
individual stages of the evolution and will now be discussed in turn.
The main differences encountered here
have to do with gauge and excision.

\subsection{Pre-Merger Evolution}
\label{sec: pre-merger evolution}

Pre-merger, our spacetime contains two apparent horizons.
Within each apparent horizon, there is a curvature singularity, hidden from us
by singularity excision. As the holes approach each other, the excised
regions -- or \emph{excision masks} -- are allowed to move also.
The new locations of the excision masks must be found by somehow tracking the
locations of the black hole singularities. One possibility is to track the
singularities via an apparent horizon finder; we have found, however, that
a simple Gaussian fit to the shape of the trace of the extrinsic curvature
is equally effective for this purpose.
As the excision regions move,
grid points that were previously excised become part of the
computational domain. These uncovered points
must be supplied with valid data for
all fields; this ``re-population'' of points is achieved through extrapolation
onto the mask surface along coordinate normals to the surface. Both the
Gaussian mask tracking and the re-population procedure have been treated in
detail in Papers 1 and 2.

Besides the above, the current problem of evolving binary data raises new
challenges. Primary among these is the question of gauge conditions.
While substantial progress has been made in the recent past in
deriving successful gauge conditions for evolutions that anchor black holes
to a fixed coordinate location (see
e.\,g.\,\cite{Bona1995, Alcubierre2001b, Alcubierre2001, Yo2002,
Alcubierre2003, Alcubierre2003c, Bruegmann2004}) the question
as to suitable gauge conditions for strongly time varying scenarios,
such as spacetimes with moving black holes, remains largely unanswered.

In view of the successful simulations of single moving
black holes in Paper 2,
we have decided to pursue a similar approach for the binary system at hand,
namely analytic  shift and densitized lapse.
A key benefit of using a densitized version of the lapse is that it
facilitates an algebraic specification of the slicing condition,
in terms of an analytic function of the spacetime coordinates,
while satisfying criteria for strong and symmetric
hyperbolicity as derived in \cite{Gundlach2004}
or \cite{Sarbach2002}.
Using a densitized lapse thus leads to a well-posed Cauchy problem
as defined in \cite{Nagy2004}.
In contrast to the single black hole simulations of Paper 2,
where algebraic forms of the shift and densitized lapse were available
from the analytic single black hole solution, gauge conditions
for our head-on collisions,
in which the black holes exhibit changes in their coordinate location,
require a suitable guess.
We must emphasize that any gauge condition we introduce should not in principle
affect the physical content in the spacetime. In practice, we
specify gauge conditions that facilitate the merger and achieve long-term-stable
evolutions.

We require that the
lapse function and shift vector resemble those
of a single, boosted Kerr-Schild black hole
in the vicinity of each singularity, namely
Eqs.~(\ref{ID:KS_lapse}) and (\ref{ID:KS_shift}) respectively.
Using these single, boosted black hole expressions, we
construct the following gauge conditions:
\begin{eqnarray}
\label{eq:ID_BBH_gauge1}
    Q       & = & \gamma^{-1/2} \left( \,^A \alpha^{-2}
                  + \,^B \alpha^{-2} - 1\right) ^{-\frac{1}{2}}, \\
                  \label{eq:ID_BBH_gauge2}
    \beta^i & = & \gamma^{i j} \left( \,^A \beta_j + \,^B \beta_j \right).
\end{eqnarray}
Here $Q$ represents the densitized lapse, $\gamma_{ij}$
is the analytic three-metric given in Eq.\,(\ref{ID:HuMaSh_g}),
$\gamma$ its determinant and $\,^A \alpha$, $\,^B \alpha$,
$\,^A \beta_j$, $\,^B \beta_j$ are the analytic lapse and
covariant shift
of holes $A$ and $B$.
It is not difficult to show that the expressions
(\ref{eq:ID_BBH_gauge1}) and (\ref{eq:ID_BBH_gauge2}) have a
close limit. That is, at zero separation and boost velocity, they
coincide with the densitized lapse and shift
of a single, non-rotating, non-boosted
Kerr-Schild hole of mass $M = 2 m$.

It is important to emphasize that the gauge conditions
(\ref{eq:ID_BBH_gauge1}) and (\ref{eq:ID_BBH_gauge2}) require information
about the coordinate location of the black hole singularities
and boost velocities.
Here, black hole coordinate location is not meant to be a formal
statement about the precise location of the physical singularities,
but it is merely used to represent the point
with respect to which (\ref{eq:ID_BBH_gauge1}) and (\ref{eq:ID_BBH_gauge2})
are evaluated. Therefore, in order to use these gauge conditions,
there remains the task of
finding suitable trajectories for the black holes.
We have found the following ``Newtonian" trajectories
adequate for this purpose:
\begin{equation}
  \vec{r}(t) = \vec{r}_0 + \vec{v}_0 t + \frac{1}{2} \vec{a}_0 t^2\,.
\label{eq:HO_pNewt_traj}
\end{equation}
Here $\vec{r}_0$ and $\vec{v}_0$ are the initial singularity position and
boost velocity, respectively. 
The value for $\vec{a}_0$ is obtained iteratively using a
small number of preliminary runs
to ensure the largest gradients of the gauge
fields remain well covered by the excision mask.
During the evolution, the boost velocity needed to compute
(\ref{eq:ID_BBH_gauge1}) and (\ref{eq:ID_BBH_gauge2}) is obtained
simply from $\vec{v} = d\vec{r}/dt$. 

\subsection{The Merger of Excision Masks}
\label{sec: merger}

The key point during the merger stage is, once a common apparent horizon
has formed and been identified, to replace the individual
spherical excision masks with a single spherical one.
The size of this new excision region is determined by the
minimum sphere containing the original excision masks.
However, special care must be taken with regard to
the timing of this replacement. One must check that the
new spherical mask is fully contained within the apparent horizon.
Typically, when the common horizon forms, it has an elongated or prolate shape.
Therefore, a delay in merging the mask allows the common apparent horizon
to become more spherical, thus facilitating the creation of
a spherical merged mask. On the other hand,
there is a constraint on how long this delay can be.
Because of the extrapolations involved in the excision,
one cannot delay the merger of the masks until their surfaces
touch each other.
Even if the common apparent horizon is not completely spherical,
the masks are merged if there are not enough grid-points between the
individual masks to carry out excision extrapolations. Of course, in this
case one still has
to continue to satisfy the requirement that the merged mask is contained
within the apparent horizon. For
the head-on collisions under consideration, we find that
the common apparent horizon
becomes almost spherical before the separation of individual masks decreases
to the point of
preventing excision extrapolations.

Resolution also plays a factor during the mask merger.
If the grid resolution in the vicinity of the black holes is too
coarse, the minimum coordinate separation of the individual masks
required by excision
extrapolations will be too large. In consequence, it
becomes more difficult to find a spherical merged mask that comfortably
fits within the common apparent horizon.
For the typical
resolution used in our simulations, $h = 0.135\,M$,
two single-hole masks of radius $0.405\,M$ require a merged mask with radius
of $1.1475\,M$. That is, single-hole masks with 40.5\,\% of the
single-hole horizon radius mean the new excision radius is approximately
57\,\% of the common horizon radius.

\subsection{Post-Merger Ring-Down}
\label{sec: post-merger}

After the excision mask merger, the key change concerns the
gauge conditions. Post-merger, the spacetime effectively represents
a single ringing black hole, and the problem at hand is similar to
that of evolving a single stationary hole. In Papers 1 and 2, we have
demonstrated how this scenario can be simulated with long-term stability
using either live gauge conditions or algebraic gauge using a
densitized lapse. The same options apply to the post-merger stage
of a black hole collision, and we have found that either approach
provides long-term stability. For the
extraction of wave forms, we find a similar situation: live and densitized
algebraic gauge conditions
generate waveforms that differ only within numerical errors. For simplicity,
we use densitized algebraic gauge conditions (see Paper 1 and 2 for details).

There remains the question of how the
transition from the superposed gauge (\ref{eq:ID_BBH_gauge1})
and (\ref{eq:ID_BBH_gauge2})
to a single black hole gauge takes place.
As mentioned before, the gauge conditions (\ref{eq:ID_BBH_gauge1})
and (\ref{eq:ID_BBH_gauge2}) have the correct close limit
(i.e. vanishing separation and boost velocities).
Starting at the time when the masks are merged, we match the
current ``black hole positions" given by (\ref{eq:HO_pNewt_traj})
to decelerating trajectories given by
sixth-order polynomials in time. These new trajectory functions
are such that
the coordinate separation and
boost velocities vanish within a time interval (typically $10\,M$).
The order of the polynomial is fixed by requiring
continuity of the boost velocity and acceleration.

\section{Results}
\label{sec: results}

The spacetime of head-on collisions of non-spinning black holes
is by construction axisymmetric. In the equal-mass case there
is a further reflection symmetry across the plane perpendicular to
the collision axis halfway between the holes.
Although the simulations
in this work are intrinsically 3D, we take advantage of
these symmetries and evolve only one
octant of the numerical domain.
One of the primary goals of our work is to perform 3D-simulations
of head-on
collisions with initial separations never tried before.
However, determining what constitutes a large separation has
the following complication.
Previous work on head-on collisions has used exclusively
Misner or Brill-Lindquist initial data, as for example
the recent long-term-stable simulations of Alcubierre
et al.\,\cite{Alcubierre2003b,
Alcubierre2004}.
To our knowledge, the only evolutions of binary Kerr-Schild (HuMaSh) data is the
grazing collision of Brandt et al.\,\cite{Brandt2000}, where the
black holes had an initial coordinate separation of $5\,M$.

We must emphasize that there exists no unambiguous definition of
a black hole separation; also,
given that we are using HuMaSh initial data,
direct comparison of our initial black hole coordinate separations
with those from Misner or Brill-Lindquist initial data is not correct.
An operational measure of
higher physical significance is given by
the horizon-to-horizon proper separation. That is, the
proper 3D distance horizon-to-horizon
computed along the collision axis within the hypersurface where
the initial data resides.
One should bear in mind that this proper separation is
slicing-dependent. Additionally, our proper distance
separation suffers from the
effects due to constraint violations intrinsic to the HuMaSh data.

Table \ref{tab: trajectories} gives the black hole coordinate ($D$)
and proper ($L$) separations for the
two sets of runs (labeled $D06$ and $D10$) performed in this work.
In addition, the table gives the initial boost velocity
($v_0$) as well as the ``acceleration" parameter ($a_0$) used
to compute the trajectories required to construct
the gauge variables [cf.\,Eq.~(\ref{eq:HO_pNewt_traj})].
As a reference, the ``Cook-Baumgarte'' ISCO
\cite{Baumgarte2000, Cook1994} corresponds to
a proper separation between $4.8$ and $4.88\,M$.
Similarly, the orbit simulation by
Br\"ugmann et al.\,\cite{Bruegmann2004}
started from a proper separation of about $9\,M$.

We evolve the initial data
on a set of four nested refinement levels, with resolution and
extent as given in Table \ref{tab: grid}.
Except for the outermost level, grid spacing and extent always increase
by a factor of two. We have found that
a grid spacing significantly larger than 
$1.08\,M$ in the outer regions gives rise to substantial reflections
of gravitational waves at the refinement boundary. For this reason we have
pushed the outermost level to $140.4\,M$ without adding a further coarser
level.

\begin{table}
  \caption{\label{tab: trajectories}
Setup parameters for the runs. 
$D$ and $L$ are the coordinate and proper initial separations.
$v_0$ is the boost velocity and $a_0$ the ``acceleration" parameter
to compute the gauge trajectories.}
  \begin{ruledtabular}
  \begin{tabular}{l|rrrr}
    Run & $D\,[M]$ & $L\,[M]$ & $v_{0}$ & $a_{0}\,[M^{-1}]$ \\
    \hline
    $D06$ &  $6$ &  $5.49$  & $0.289$  & $0.040$ \\
    $D10$ & $10$ & $10.06$  & $0.270$  & $0.014$ \\
  \end{tabular}
  \end{ruledtabular}
\end{table}
\begin{table}
  \caption{\label{tab: grid}
           Grid parameters of the refinement levels used in the simulations.
           The number of grid points $N$ includes ghost, symmetry
           and outer boundary zones. $r_{\rm max}$ is the extent in the
           $x$, $y$ and $z$-directions.}
  \begin{ruledtabular}
  \begin{tabular}{c|ccc}
    ref-level & $h\,\,[M]$ & $N$ & $r_{\rm max}\,\,[M]$ \\
    \hline
    1 & $0.135$ & $67^3$ & $8.505$ \\
    2 & $0.27$  & $67^3$ & $17.01$ \\
    3 & $0.54$  & $67^3$ & $34.02$ \\
    4 & $1.08$  & $134^3$& $140.4$ \\
  \end{tabular}
  \end{ruledtabular}
\end{table}

\begin{figure}
  \begin{center}
    \includegraphics[width=220pt, angle=0]{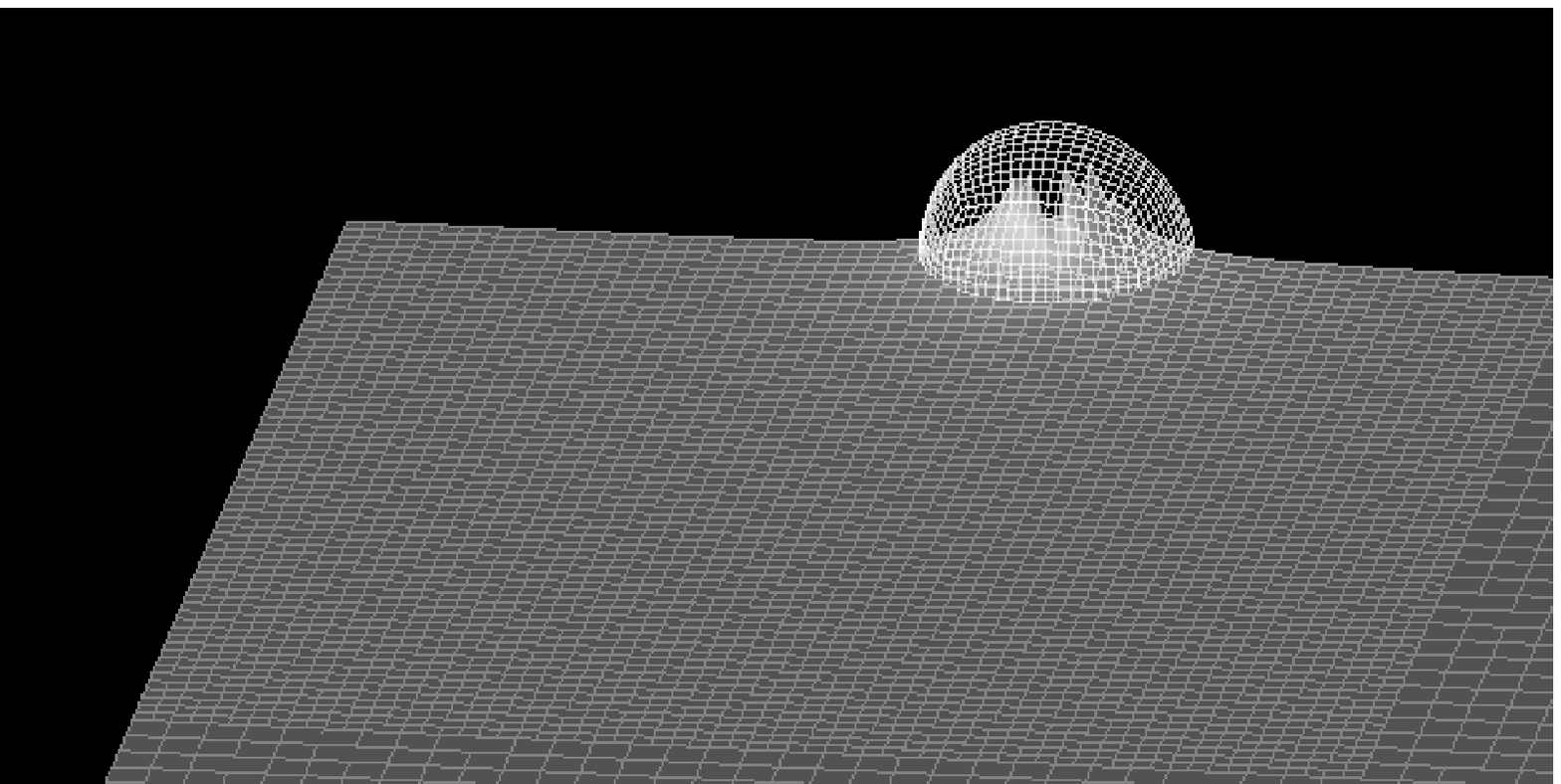}
    \includegraphics[width=220pt, angle=0]{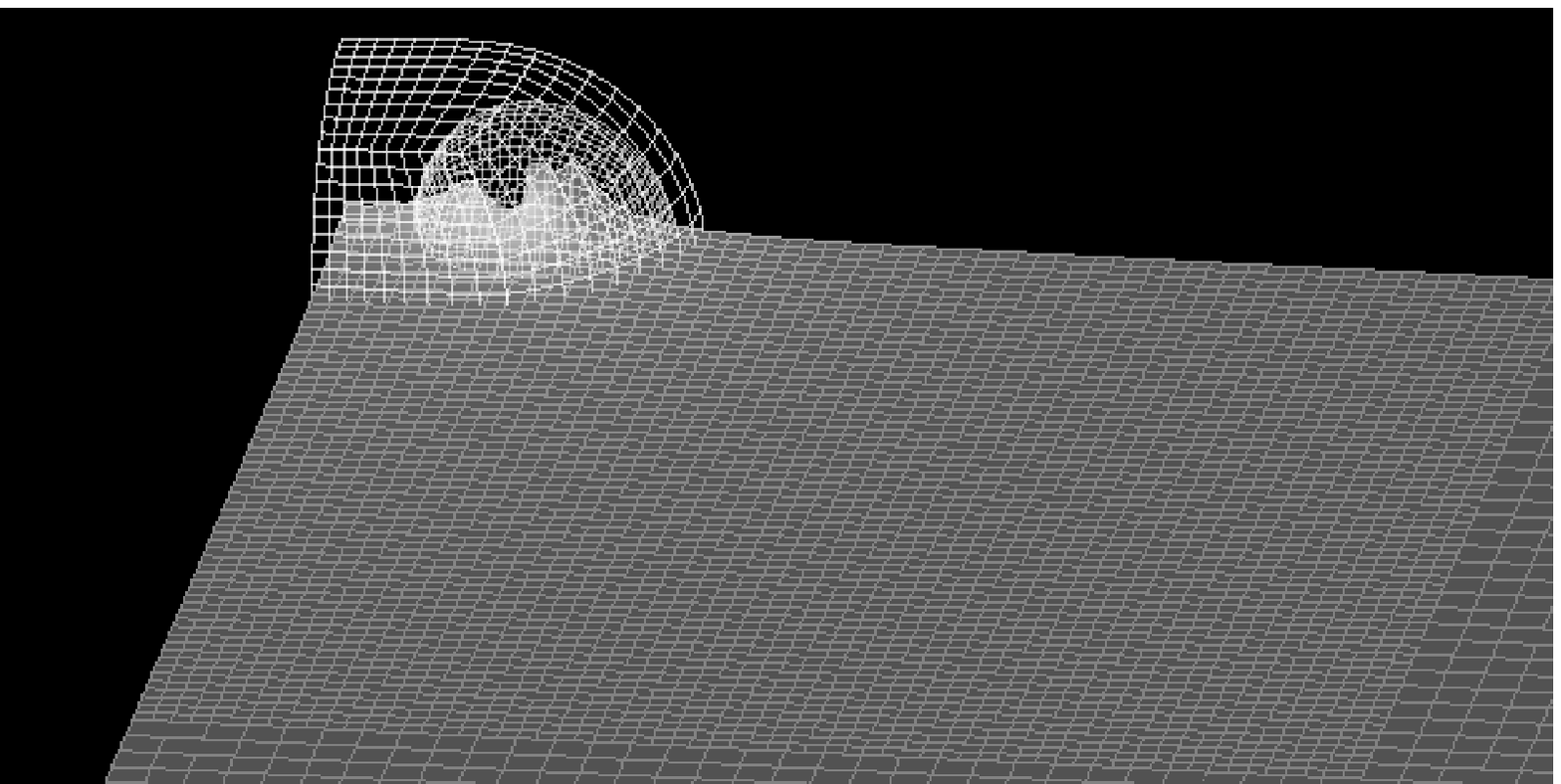}
    \includegraphics[width=220pt, angle=0]{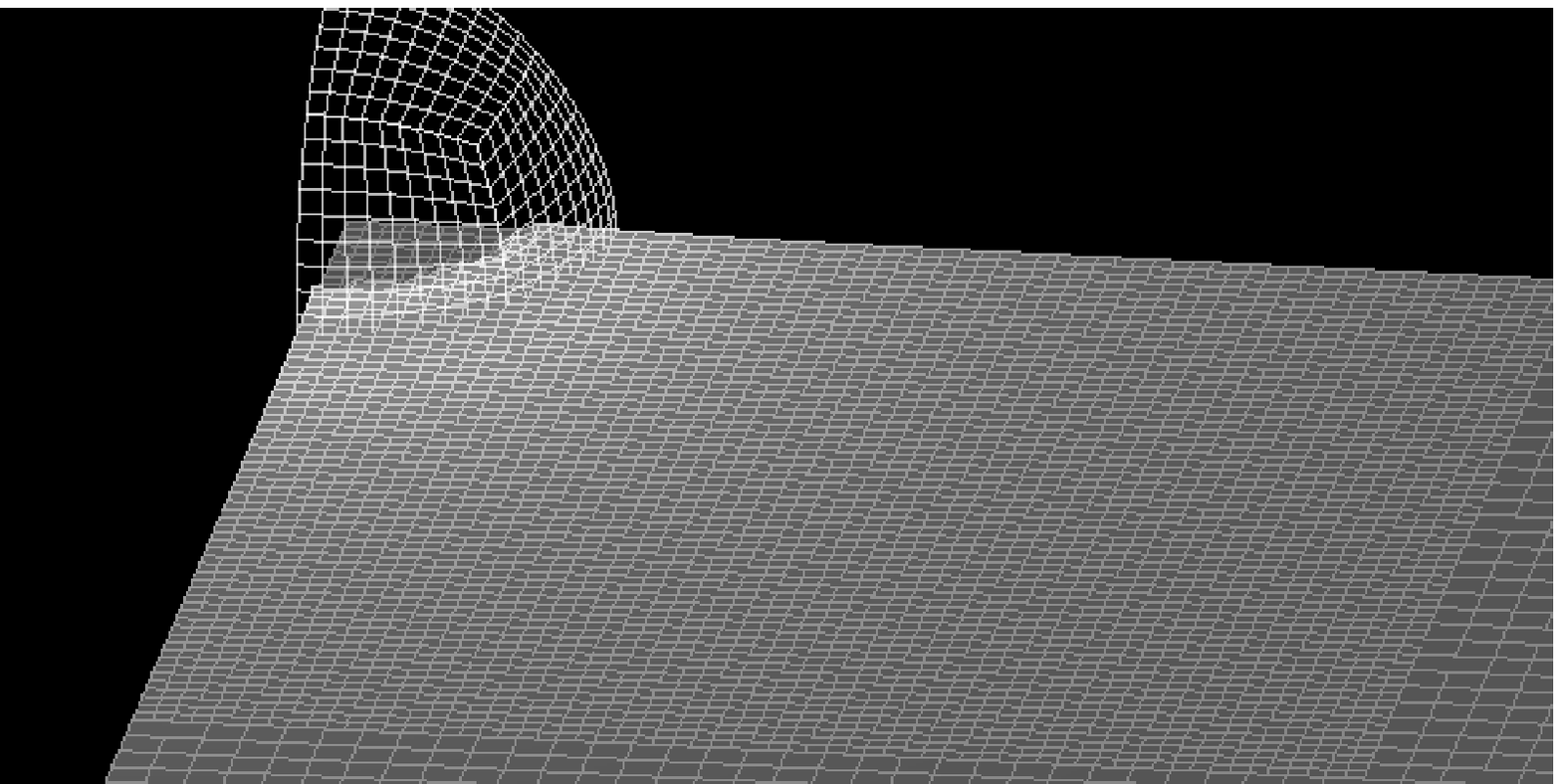}
    \caption{Snapshots of the $D10$ run
             taken at times $t=0$, $10.8$ and $14\,M$.
             From top to bottom the panels show the individual holes
             with distinct apparent horizons (only one hole is visible
             due to the octant symmetry), a common apparent horizon
             forming around the individual trapped surfaces and a single
             apparent horizon surrounding a single excision region.}
    \label{fig: time_evolution}
  \end{center}
\end{figure}

Using the setup described above, we obtain long-term-stable evolutions.
The three simulation stages
described in Sec.\,\ref{sec: numerical evolution} are illustrated in
Fig.\,\ref{fig: time_evolution}. This figure contains snapshots of the $D10$ run
at times $t=0$, $10.8\,M$ and $14\,M$, including the 
marginally trapped surfaces as calculated by Thornburg's {\sc AHFinderDirect} code
\cite{Thornburg1996, Thornburg2004}.
The upper panel shows the initial configuration with
holes present at $z=\pm 5\,M$.
As mentioned above the computational domain only includes the hole
at $5\,M$ due to symmetry.
During this pre-merger stage each hole is surrounded by its own apparent
horizon (AH). In the middle panel, the holes are sufficiently close
to one another that a common AH has formed.
However, as mentioned before, it is not possible yet
to switch to a single excision region because of the elongated shape
of the common apparent horizon.
The lower panel shows the beginning of the post-merger stage.

\subsection{Waveforms}

We extract waveforms at radii $12.5$, $20$, $25$ and $30\,M$
using the Zerilli-Moncrief formalism; details are given
in App.~\ref{sec: Zerilli}.
In principle, it is desirable to extract gravitational waves at
large radii so as to ensure that the assumptions inherent to the
Zerilli-Moncrief extraction are satisfied.
We observe two problems, however, which arise out of choosing
excessively large radii. First,
outer boundary effects require less time to propagate inward and reach
the extraction radius. Second, the actual wave signal contained in the
metric and curvature variables is weaker at larger radii and thus
requires higher accuracy in the numerical evolution. In combination with the
increasingly coarse resolution used in outer levels,
this gives rise to a degree of
distortion and noise in the extracted Zerilli function which is not present
at smaller radii. We find the values used above to be a reasonable compromise
for the post-merger stage of the head-on collision.
\begin{figure}
  \begin{center}
    \includegraphics[height=270pt, angle=-90]{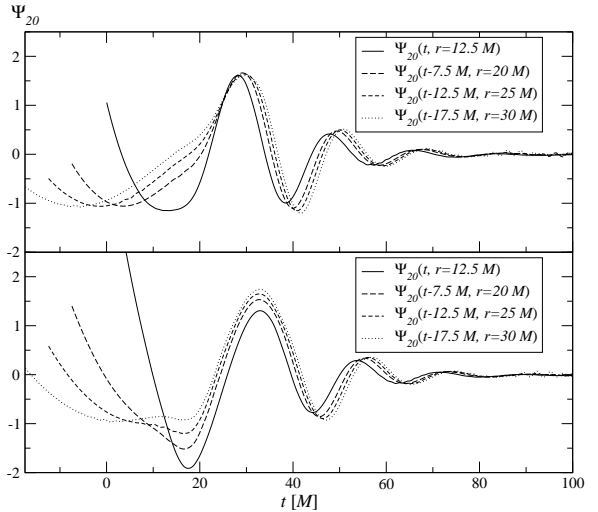}
    \caption{The $\ell=2, m=0$ Zerilli function extracted at different radii
             for the $D06$ (upper panel) and $D10$ (lower panel) runs.
             The curves have been shifted in time by the difference in
             propagation time of the wave signal between the different radii.}
    \label{fig: zerilli06_10}
  \end{center}
\end{figure}
\begin{figure}
  \begin{center}
    \includegraphics[height=270pt, angle=-90]{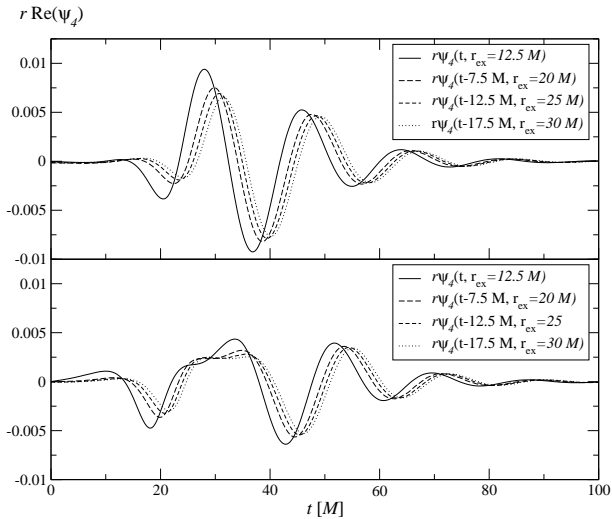}
    \caption{The $\ell=2$, $m=0$ component of the real part of the
             Newman-Penrose scalar $\psi_4$
             extracted at different radii
             for the $D06$ (upper panel) and $D10$ (lower panel) runs.
             The curves have been shifted in time by the difference in
             propagation time of the wave signal between the different radii.}
    \label{fig: weyl}
  \end{center}
\end{figure}
In Fig.\,\ref{fig: zerilli06_10} 
we show the resulting $\ell=2, m=0$
waveforms, each shifted in time by the amount required by the signal
to travel from $r=12.5\,M$ to the corresponding extraction radius.
All curves in the figure
show a significant initial signal up to about $t\approx 10\,M$.
We trace the origin of this signal to the fact that the two black holes are far
apart in the early stages of the simulation and thus the assumption
underlying the wave extraction technique, namely that the spacetime
is that of a single black hole plus a perturbation, is not satisfied
with sufficient accuracy during the pre-merger phase.
This viewpoint is strengthened by the observation that the spurious
signal is stronger for the larger black hole separation of
case $D10$ in the lower panel of the figure.
We further note in this context
that the merger of the black holes
occurs at $t=5.4\,M$ and $11.75\,M$ respectively for cases $D06$ and $D10$.
It is at about these times that the shifted waveforms extracted at different
radii start showing better agreement with each other in the figures,
while the spurious signal present at earlier times converges away
at larger extraction radii.

Because the Zerilli potential becomes zero in the limit of infinite
extraction radii,
one would expect the curves in the individual panels
of Fig.\,\ref{fig: zerilli06_10}
to overlap exactly in this limit.
The figure shows that
this criterion is satisfied with increasing accuracy at larger radii. Still
there remain small effects of back-scattering of the wave signal at the
finite radii used in the extraction. The figure also reveals some noise
developing at late times at larger radii; an analysis
of the timing of the noise demonstrates that these are boundary effects
propagating inward. Below, we will see that these effects limit the
convergence properties of our code and thus the reliability of the
results after the onset of noise in the waveforms.

An alternative to the Zerilli-Moncrief approach to wave extraction
is provided by the Newman-Penrose scalar $\psi_4$.
Given the Weyl tensor $C_{a b c d}$, and an appropriately
chosen null tetrad $\{ l, n, m, \bar{m} \}$,
$\psi_4 \equiv C_{a b c d} \, \bar{m}^a \, n^b \,\bar{m}^c \, n^d$
represents a measure of the outgoing gravitational wave content.
We calculate $C_{a b c d}$
using the ``electric-magnetic'' decomposition
\cite{Smarr1975} as described in \cite{Gunnarsen1995}.
For spacetimes perturbatively close to Kerr,
the appropriate choice of null tetrad is the Kinnersley tetrad
\cite{Kinnersley1969}
In practice, the exact Kinnersley tetrad is not
available in a numerical simulation.
Even though recent work has suggested that
it may be possible to identify unambiguously the Kinnersley null directions
of the numerical spacetime \cite{Beetle2004},
the resulting Weyl scalars would still differ from their Kinnersley
values in polarization and scaling.
For simplicity we use instead a symmetric null tetrad formed from an
orthonormalized spherical coordinate tetrad:
\begin{eqnarray}
  &l^a \equiv \frac{1}{\sqrt{2}} \, ( u^a + e_r^a ) \, , \, &n^a \equiv
    \frac{1}{\sqrt{2}} \, ( u^a - e_r^a ) \, , \nonumber \\
    &m^a \equiv \frac{1}{\sqrt{2}}
    \, ( e_{\theta}^a + i \, e_{\phi}^a )\, ,
    &\bar{m}^a \equiv \frac{1}{\sqrt{2}}( e_{\theta}^a - i \, e_{\phi}^a ).
\end{eqnarray}
Here $u$ is the future-pointing unit hypersurface normal and
the coordinate triad vectors were orthonormalized via a Gram-Schmidt
process in the order $e_r$, $e_{\theta}$, $e_{\phi}$.
A symmetric tetrad of this kind (though possibly orthonormalized in a different
order) is a
standard choice in approximately spherically symmetric spacetimes
\cite{Smarr1979,Zlochower2005,Fiske2005}, and is the starting point of the
Lazarus procedure \cite{Baker2002}.
While the numerical three-metric is used for orthonormalization,
the tetrad requires no explicit knowledge of a background spacetime,
and so the resulting
expressions do not involve parameters such as a background mass and angular
momentum.
At large extraction radii,
the leading difference between these tetrad components will be an
overall (radius-dependent) scaling factor, which is not relevant for
our purposes. Additionally, any coordinate distortion should be a first-
order effect, and should not influence the waveforms to leading order.
That said, we cannot quantify the relative strength at early times of the
assumptions behind Weyl extraction of this type and Zerilli extraction as
described earlier.

In Fig.\,\ref{fig: weyl} we plot the $\ell=2$, $m=0$ quadrupole moment
of the real part of $r\psi_4$. The application of the factor $r$ is
convenient because the resulting function will propagate in the
limit of large radii in analogy to the Zerilli function.
While the
waveforms in Figs.\,\ref{fig: zerilli06_10}, \ref{fig: weyl} exhibit
good agreement during the merger and ring-down phase, differences are
apparent during the early in-fall stage. In particular the Newman-Penrose
waveform is less severely affected by the deviations from an approximately
spherically symmetric spacetime. While a detailed comparison between
the two methods of gravitational wave extraction is beyond the scope of
this work, it will be interesting to analyze the observed discrepancies
in more depth in
future work. The same applies to the distortion visible in the
Newman-Penrose waveform for the simulation $D10$ around $t=30\,M$.

In the remainder of this work we will restrict our analysis to
the waveforms of Fig.\,\ref{fig: zerilli06_10}
obtained with the Zerilli-Moncrief formalism.
In order to estimate the quality of the waveforms, next we
investigate their convergence properties.

\subsection{Convergence}
\label{sec: convergence}

The discretization of the BSSN-evolution equations in the Maya code is
implemented with second-order accuracy in the grid spacing $h$.
The wave extraction, as described in Appendix~\ref{sec: Zerilli},
involves numerical approximations of higher-order
accuracy, such as the interpolation
of grid functions onto the sphere of extraction and the evaluation
of the integrals over the spheres.
The numerical error in the wave functions should therefore be dominated
by the second-order accuracy of the evolution of the field equations,
and we expect the wave forms to exhibit second-order convergence.

In order to verify this numerically, we focus our attention
on the head-on collision
case $D06$ and carry out two additional runs.
Compared with the resolutions listed in
Table \ref{tab: grid}, the two additional
runs are finer by factors of $1/1.5$ and $1/1.5^2$.
On the finest grid for each of the three runs, this leads to
resolutions $h_1=0.135\,M$, $h_2=0.09\,M$ and $h_3=0.06\,M$, respectively.
Because the coarsest mesh in $D06$ has an extent of $140.2\,M$,
implementing the $1/1.5$ and $1/1.5^2$ refinements
far exceeds the available memory of our hardware resources.
For this reason, instead of using mesh-refinements with the sizes given
in Table \ref{tab: grid}, we have performed the convergence analysis using
for each run four refinement levels of extent $r_{\rm max}=5.4$, $10.8$,
$21.6$ and $43.2\,M$. The key difference compared with
the simulations discussed
above is that outer boundary effects will reach the extraction
radii at earlier times. To mitigate this effect we
extract the Zerilli function at the smaller radius $r_{\rm ex}=7.5\,M$
for this analysis.
\begin{figure}
  \begin{center}
    \includegraphics[height=270pt, angle=-90]{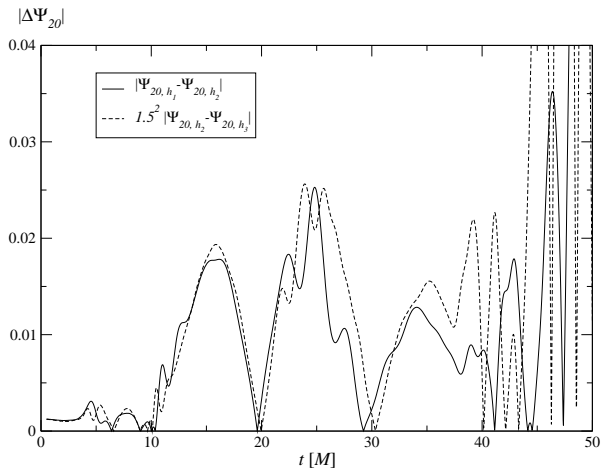}
    \caption{Three-level convergence analysis for the $\ell=2$, $m=0$
             Zerilli function extracted at $r=7.5\,M$.}
    \label{fig: conv_abs}
  \end{center}
\end{figure}

In Fig.\,\ref{fig: conv_abs} we show the difference in the $\ell=2$, $m=0$
Zerilli function obtained for the three different resolutions. The difference
in $\Psi_{20}$ between the medium and high resolution runs has been
amplified by the factor $1.5^2$ expected for second-order convergence.
The resulting curve agrees reasonably well with the difference obtained from
using the low and medium resolution until about $35\,M$, indicating
second-order convergence. After that time, noise originating from the
outer boundary at $43.2\,M$ has reached the extraction radius at
$7.5\,M$ and spoils the convergence properties.
This result is confirmed by analysis at larger radii where second-order
convergence is observed for correspondingly shorter periods of time.
We conclude that our code facilitates second-order convergent wave
extraction until outer boundary effects have had the time to propagate
inward toward the radius of wave extraction. For the simulations
presented in the previous subsection, this result implies reliable
waveforms up to a time of about $100-125\,M$ depending on the extraction
radius.

In view of this result it appears desirable to achieve more
satisfactory boundary treatments. Various alternatives have been
suggested to this effect in the past. Conceptually the most elegant
approach consists in the incorporation of null-infinity in the numerical
domain via the so-called {\em Cauchy-characteristic matching}. The key
feature of this technique is the matching of
a ``3+1'' formulation in the interior regions containing the
strong-field sources to a characteristic formulation of the
exterior spacetime (see e.\,g.\,\cite{Bishop1996,dInverno1996,Winicour2001}).
This facilitates a straightforward
compactification of the spacetime and, thus, the implementation of
exact boundary conditions.
An alternative matching technique has been suggested in \cite{Abrahams1998}.
Here the weak-field exterior spacetime is described
in terms of non-spherical perturbations of a Schwarzschild background.
A different strategy without matching involves the
development of alternative boundary conditions which by construction
preserve the constraint equations (see e.\,g.\,\cite{Frittelli2004,
Gundlach2004, Beyer2004}).
While the derivation of such
boundary conditions has largely been motivated by stability issues, it would
be interesting to see their effect on
spurious reflections as observed in our simulations.
The implementation of either of these alternative boundary treatments
represents a considerable challenge, however,
and is beyond the scope
of this work. Instead we reduce the impact of boundary effects by
placing the outer boundaries far from the strong field sources.

Finally we emphasize that
in spite of the accuracy limitations arising from our implementation
of the boundary conditions,
we have not observed any sign of instabilities:
all simulations presented in this work have continued for many hundreds or
even thousands of $M$ gradually approaching a stationary spacetime containing
a single black hole.

\subsection{Quasi-normal Ringing, Masses and Radiated Energy}

\begin{table}
  \caption{\label{tab: frequencies}
           Quasi-normal ringing damping $\sigma$ and
           oscillation frequency $\omega$
           measured from the extracted waveforms.
           The final column gives the percentage of the
           initial ADM mass radiated away in the form of
           gravitational waves as derived from
           Eq.\,\ref{eq: power}}
  \begin{ruledtabular}
  \begin{tabular}{cc|c|c|c}
  run & $r_{\rm ex}/M$ & $\sigma M_{\rm ADM}$ & $\omega M_{\rm ADM}$ & $E_{\rm rad}\,[\%]$ \\
  \hline
  $D06$ & $12.5$ & $0.0917$ & $0.365$ & $0.20$ \\
  $D06$ & $20$   & $0.0921$ & $0.362$ & $0.15$ \\
  $D06$ & $25$   & $0.0923$ & $0.363$ & $0.14$ \\
  $D06$ & $30$   & $0.0947$ & $0.366$ & $0.14$ \\
  \hline
  $D10$ & $12.5$ & $0.0973$ & $0.369$ & $0.37$ \\
  $D10$ & $20$   & $0.0963$ & $0.363$ & $0.16$ \\
  $D10$ & $25$   & $0.0960$ & $0.365$ & $0.14$ \\
  $D10$ & $30$   & $0.0976$ & $0.366$ & $0.13$ \\
  \end{tabular}
  \end{ruledtabular}
\end{table}
\begin{figure}
  \begin{center}
    \includegraphics[height=250pt, angle=-90]{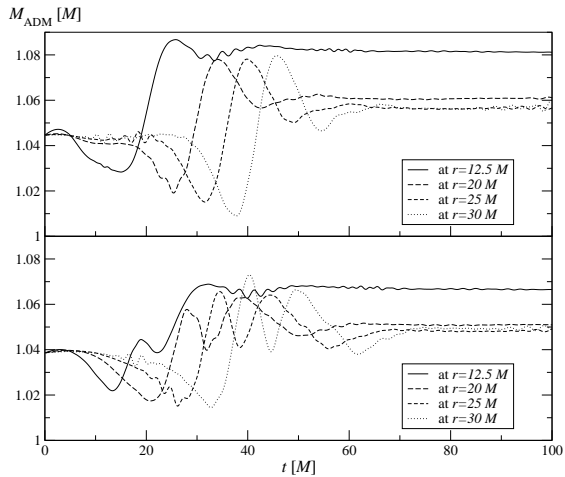}
    \caption{ADM mass as a function of time
             evaluated at various radii for the
             $D06$ (upper panel) and $D10$
             (lower panel) simulation.}
    \label{fig: adm}
  \end{center}
\end{figure}
\begin{figure}
  \begin{center}
    \includegraphics[height=250pt, angle=-90]{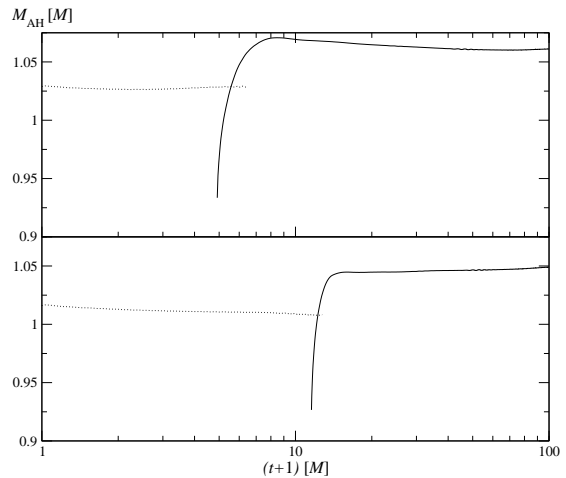}
    \caption{The apparent horizon mass as a function of time
             calculated for the $D06$
             (upper panel) and $D10$
             (lower panel) simulation.}
    \label{fig: ah_mass}
  \end{center}
\end{figure}
Previous studies
have shown that most of the gravitational radiation
of a head-on collision is emitted in the post-merger
ring-down phase, with the total radiated energy typically being less than
a percent of the total initial mass \cite{Anninos1993,
Price1994, Abrahams1994, Anninos1995d}.
Furthermore, the quasi-normal ringing of the post-merger black hole
is dominated by the $\ell=2$, $m=0$ mode with small contributions
from the $\ell=4$, $m=0$ multipole.
Our simulations represent the first head-on collisions
with wave extraction using Kerr-Schild
type data.  They are in good agreement with the
picture described above.

We first note that up to $\ell=3$ the only
multipole with a notable wave signal is the $\ell=2$, $m=0$ quadrupole shown
in Fig.\,\ref{fig: zerilli06_10}.
Second, we have performed least-square fits of the waveforms to
\begin{equation}
  f(t) = A\,e^{-\sigma t}\cos\,(\omega t - \phi_0).
\end{equation}
in order to calculate the ringing damping $\sigma$ and
oscillation frequency $\omega$.
The results are given
in Table\,\ref{tab: frequencies}.
The resulting damping and oscillation frequency values agree within
about $10\,\%$.
For comparison, the quasi-normal mode spectrum
of the Schwarzschild black hole calculated by Chandrasekhar and Detweiler
\cite{Chandrasekhar1975} predicts
$\sigma \mathcal{M} = 0.0890$, $\omega \mathcal{M} = 0.374$,
where $\mathcal{M}$ denotes
the mass associated with the
black hole background spacetime.
In Table \,\ref{tab: frequencies}, the results are
reported in units of the total initial ADM mass $M_{\rm ADM}$.
The differences between these two masses are the radiated energy
and numerical errors.

In the course of the evolution we numerically approximate the
ADM mass by evaluating the ADM integral at finite extraction radii.
The result is shown in Fig.\,\ref{fig: adm} for simulation
$D06$ in the upper panel and
$D10$ in the lower panel. The initial values agree well
with the sum $\gamma M$ of the special relativistic masses
of the two individual holes moving with
the velocities
given in Table \ref{tab: trajectories},
that is $M_{\rm ADM} = 1.045\,M$ and $1.039\,M$ respectively. During
the evolution the numerical ADM mass
remains within a few percent of the
initial value but in contrast to intuitive expectations it increases slightly
rather than decreases after the gravitational waves have passed the extraction
radius.
Even though this increase becomes less pronounced at larger extraction radii
we do not consider this calculation of the ADM mass sufficiently accurate
to support a calculation of the total radiated gravitational
wave energy from conservation arguments.

In Fig.\,\ref{fig: ah_mass}, we plot the apparent horizon masses
as calculated by {\sc AHFinderDirect} for the two cases $D06$ (upper)
and $D10$ (lower panel).
During the pre-merger stage there is no common apparent horizon
present and we plot instead the sum of the horizon masses associated
with the individual holes (dotted curves). In contrast the
solid curve represents the horizon mass of the single post-merger hole.
In the early stages after merger the horizon
mass shows a significant increase in time but quickly settles down in the
final values of $1.06\,M$ and $1.05\,M$ respectively. These values
show good agreement with the values $1.045\,M$ and $1.039\,M$ used
for the black hole background mass for the wave extraction.
Theoretically one would not expect the horizon mass to decrease
as a function of time. We attribute the decrease by about one percent
visible after the
merger in the upper panel of Fig.\,\ref{fig: ah_mass} to numerical
inaccuracies.

We estimate the total radiated energy
from the Landau-Lifshitz formula
(see e.\,g.\,\cite{Cunningham1980}) which gives the radiated power
as
\begin{equation}
  \rm P = \frac{1}{96\pi} \Psi_{,t}^2,
  \label{eq: power}
\end{equation}
where the constant factor arises from our definition of the Zerilli
function in Eq.\,(\ref{eq: zerilli}). Ideally this function needs to be
evaluated at null infinity. As this is not possible
in this type of computation,
we approximate the exact result by calculating the radiated energy at
various finite radii. For this purpose, we have evaluated the integral
of ${\rm P}$ from $t=0$ to $t=140\,M-r_{\rm ex}$; that is, we have excluded late
times when the extraction radius is causally connected to the outer boundary.
The results are shown in the last column of Table \ref{tab: frequencies}
and show the impact at small extraction radii of the spurious signal
in the Zerilli function at early times when the two holes are still
far apart from each other. This feature leads to an overestimate of the
radiated energy but quickly converges away if we extract at larger
radii. A close investigation reveals that this spurious contribution
drops to a few percent of the total radiated energy at $r_{\rm ex}=30\,M$
in the case $D10$ and even less in the case $D06$.
Within these error bounds we thus find a radiated energy of
about $0.14$ and $0.13$ percent of the total ADM mass of the system.

\section{Conclusions}

\label{sec: conclusion}

We have presented the first long-term-stable binary black
hole collisions with wave extraction
using fixed mesh-refinement, Kerr-Schild type initial
data and dynamical excision.
While the success of our dynamical excision technique has been demonstrated
in the past in the case of static and dynamic single black hole spacetimes,
the results in the present work demonstrate the applicability of our
dynamical excision to binary black hole
systems with gravitational wave generation.

Our simulations start from initial data constructed
via superposition of two single Kerr-Schild
black holes, namely HuMaSh data.
We have demonstrated by
convergence analysis of the Hamiltonian and momentum
constraints that the inherent constraint violations in the HuMaSh data
is not significant compared with the level of accuracy allowed
by current computational resources.

We have conceptually divided the evolutions
into three stages: the initial in-fall, the merger and the post-merger
ring-down of the resulting single black hole. In particular during
the first two stages, special care must be taken to ensure that the
excision region always be entirely contained inside the apparent horizon.
For this purpose, we have calculated the apparent horizon at each time
step and verified that no causal violation of the excision technique
occurs.

We have extracted gravitational waves using the Zerilli-Moncrief approach
applied to a black hole background spacetime in Kerr-Schild coordinates.
Our results confirm previous findings.
Wave emission is dominated by the quasi-normal
ringing of the single post-merger black hole. Up to $\ell=3$ the only
notable contribution is the $\ell=2$, $m=0$ quadrupole.
Comparison of the damping and oscillation frequencies with analytic
results shows good agreement.
We do not find the accuracy in calculating the ADM
or the apparent horizon mass sufficient,
however, to support a calculation of the total radiated gravitational
wave energy purely from conservation principles. Instead, we use
the Landau-Lifshitz formula to calculate the radiated energy from the
Zerilli function. We find this energy to be of the order of $0.1\,\%$
of the ADM mass.
Our results demonstrate the need to extract gravitational radiation at
sufficiently large radii in order to facilitate reasonable accuracy.
In particular, we observe a spurious wave signal in the Zerilli
function caused by a systematic deviation during the early
pre-merger stage of the numerical spacetime from a perturbed single
black hole spacetime. The wave contribution due to this spurious
feature converges away as the extraction is carried out at larger radii.

We have also calculated the Newman-Penrose scalar $\psi_4$ as an alternative
measure of the emitted gravitational radiation. While the results
qualitatively confirm those obtained using the Zerilli approach
during the merger and ring-down phase, differences are apparent
during the early fall-in phase. In particular we find the Newman-Penrose
waveforms to be less severely affected by the deviations from an
approximately spherically symmetric spacetime at early stages.
On the other hand the Newman-Penrose scalar $\psi_4$ shows
some distortion in the case of larger initial separation of the holes.
In future work we plan to investigate
in more detail the causes of these distortions as well as compare
the different wave extraction techniques and their discrepancies at early
stages.

\begin{acknowledgments}
We thank David Fiske for pointing out an error in our calculation of the
multipoles of the Newman-Penrose scalar $\psi_4$.
We further
thank Carlos F. Sopuerta for helpful discussions about the wave extraction,
Jonathan Thornburg for providing AHFinderDirect, Thomas Radke for help
with data visualization and Deirdre Shoemaker for discussions on
gauge trajectories. We also express our thanks to the Cactus team.
We acknowledge the support of the Center for Gravitational Wave
Physics funded by the National Science Foundation under Cooperative Agreement
PHY-0114375. Work partially supported by NSF grants PHY-0244788 to Penn State
University. Additionally, B.K. acknowledges the support of the Center for
Gravitational Wave Astronomy funded by NASA (NAG5-13396) and the NSF grants
PHY-0140326 and PHY-0354867. E.S. acknowledges support from the
SFB/TR 7 ``Gravitational Wave Astronomy''.
\end{acknowledgments}

\appendix
\section{Wave extraction using the Zerilli-Moncrief formalism}
\label{sec: Zerilli}

Our extraction of gravitational waves is based on the gauge invariant
perturbation formalism of Moncrief \cite{Moncrief1974}. A more detailed
description of how this facilitates the extraction of gravitational waves
from numerical simulations is given in \cite{Abrahams1990}.
Applications of this method can be found, for example, in
\cite{Anninos1995d}, \cite{Abrahams1992}, \cite{Camarda1999}
or \cite{Font2002}.

The key idea is
that at sufficiently large distance from a strong field source, the
spacetime metric $g_{\mu \nu}$
can be viewed as a spherically symmetric
background plus non-spherical perturbations which obey the
Regge-Wheeler equation (odd) and the Zerilli equation (even
parity perturbations) respectively. In our case the background
will be given by that of a single, non-rotating black hole
in Kerr-Schild coordinates. Strictly speaking, the coordinates
are ingoing-Eddington-Finkelstein coordinates, namely
Kerr-Schild coordinates in the case of a non-rotating black hole.
The symmetry properties of an equal mass head-on collision imply that
the odd perturbations vanish, so we will
restrict our attention to the even parity case.

Given the Kerr-Schild background metric
\begin{eqnarray}
  g^{\rm B}_{\mu \nu} dx^{\mu} dx^{\nu} &=& -\left( 1-\frac{2\mathcal{M}}{r}
         \right) dt^2
         + \frac{4\mathcal{M}}{r} dt\,dr \nonumber \\
      && + \left(1+\frac{2\mathcal{M}}{r} \right) dr^2 + r^2d\Omega\,,
         \label{eq: background}
\end{eqnarray}
the first-order metric perturbations, expanded in Regge-Wheeler
tensor harmonics \cite{Regge1957}, can be written as
\begin{equation}
  h_{\mu \nu} = \sum_{\ell m} h^{\ell m}_{\mu \nu},
      \label{eq: perturbations}
\end{equation}
where
\begin{widetext}
\begin{equation}
  h^{\ell m}_{\mu \nu} = \begin{pmatrix}
      H^{\ell m}_{0} Y_{\ell m} & H^{\ell m}_1 Y_{\ell m}
      & h^{\ell m}_0 \partial_{\theta}Y_{\ell m} & h^{\ell m}_0
      \partial_{\phi} Y_{\ell m} \\[10pt]
    {\rm Sym} & H^{\ell m}_2 Y_{\ell m} & h_1^{\ell m}\partial_{\theta}
      Y_{\ell m} & h_1^{\ell m} \partial_{\phi}Y_{\ell m} \\[10pt]
    {\rm Sym} & {\rm Sym} & r^2\left( K^{\ell m}
      + G^{\ell m} \partial_{\theta \theta}
      \right) Y_{\ell m} & r^2G^{\ell m}\left( \partial_{\theta} - \cot \theta
      \right) \partial_{\phi}Y_{\ell m} \\[10pt]
    {\rm Sym} & {\rm Sym} & {\rm Sym} & r^2\sin^2\theta\left[ K^{\ell m}
      +G^{\ell m} \left(\cot \theta\, \partial_{\theta} +
      \sin^{-2}\theta\,  \partial_{\phi \phi}
      \right) \right] Y_{\ell m}
  \end{pmatrix}.
\end{equation}
\end{widetext}

Here the spherical harmonics $Y_{\ell m}(\theta,\phi)$ form a complete
orthonormal system and are eigenvectors of the operator
$-\left[ \partial_{\theta \theta} + \cot \theta \partial_{\theta}
+ \sin^{-2}\theta \partial_{\phi \phi} \right]$ with eigenvalue
$\ell(\ell+1)$.
A straightforward calculation shows that the following linear combinations
of the perturbation functions are invariant under first-order coordinate
transformations
\begin{eqnarray}
  q_1^{\ell m} &=& \frac{2\mathcal{M}r^2}{r-2\mathcal{M}}\partial_t K^{\ell m}
                   +r\frac{r-3\mathcal{M}}{r-2\mathcal{M}} K^{\ell m}
                   - \frac{4\mathcal{M}^2}{r-2\mathcal{M}} H^{\ell m}_0
                   \nonumber \\
                &&  + r^2\partial_{r} K^{\ell m} -(r-2\mathcal{M}) H^{\ell m}_2
                   - 4\mathcal{M}H^{\ell m}_1, \nonumber \\[10pt]
  q_2^{\ell m} &=& 2h^{\ell m}_1 + \frac{4\mathcal{M}}{r-2\mathcal{M}}
                   h^{\ell m}_0
                   -r^2 \partial_r G^{\ell m}
                   -\frac{2\mathcal{M}r^2}{r-2\mathcal{M}}
                   \partial_t G^{\ell m} \nonumber \\
                && - \frac{r^2}{r-2\mathcal{M}} K^{\ell m}. \label{eq: q1q2}
\end{eqnarray}
The linearized Einstein field equations
are then equivalent to two equations
expressed in terms of these gauge invariant variables. The first of these
equations is the constraint
\begin{equation}
  \left( \partial_r+2\frac{\mathcal{M}}{r-2\mathcal{M}} \partial_t\right)
  \left[ \left(1-\frac{2\mathcal{M}}{r} \right)(2q_1+\ell(\ell+1)q_2)
  \right] = 0.
\end{equation}
The second equation is more conveniently written in terms of the
Zerilli function defined in terms of the gauge invariant variables by
\begin{equation}
  \Psi_{\ell m} = \frac{-2(r-2\mathcal{M})}{[\ell(\ell+1)-2]r+6\mathcal{M}}
                  [2q^{\ell m}_1
                  + \ell(\ell+1)q^{\ell m}_2], \label{eq: zerilli}.
\end{equation}
This function represents the dynamical degree of freedom of the
polar perturbations of the black hole and obeys the second independent
combination of the Einstein equations
\begin{widetext}
\begin{equation}
  \left(1+\frac{2\mathcal{M}}{r} \right) \partial_{tt} \Psi_{\ell m}
       - 4\frac{\mathcal{M}}{r} \partial_{tr} \Psi_{\ell m} - \left(
       1-\frac{2\mathcal{M}}{r} \right) \partial_{rr}\Psi_{\ell m}
       + \frac{2\mathcal{M}}{r^2}(\partial_t \Psi_{\ell m}
       - \partial_r \Psi_{\ell m}) + V_{\ell} \Psi_{\ell m} = 0,
       \label{eq: zerilli_KS}
\end{equation}
with the potential
\begin{equation}
  V_{\ell} = \frac{\ell(\ell+1)[\ell(\ell+1)-2]^2r^3
      + 6[\ell(\ell+1)-2]^2\mathcal{M}r^2 + 36 [\ell(\ell+1)-2]
      \mathcal{M}^2r+72\mathcal{M}^3}
      {r^3\{[\ell(\ell+1)-2]r+6\mathcal{M}\}^2}.
\end{equation}
\end{widetext}
This is the well known
Zerilli equation written in Kerr-Schild background coordinates
and represents a special case of the
derivation of the Zerilli equation for arbitrary spherically
symmetric spacetimes in \cite{Sarbach2001} based on the gauge-invariant
formalism of Gerlach and Sengupta \cite{Gerlach1978}.
The more familiar form of the Zerilli equation
(see e.\,g.\,the recent review \cite{Nagar2005})
\begin{equation}
  \partial_{t_*t_*} \Psi_{\ell m} - \partial_{r_*r_*}\Psi_{\ell m}
       + \bar{V}_{\ell}\Psi_{\ell m} = 0
\end{equation}
using Schwarzschild time $t_*$ and the
tortoise coordinate $r_*$ is directly obtained from the
coordinate transformation
\begin{eqnarray}
  t_* &=& t-2\mathcal{M} \ln \frac{r-2\mathcal{M}}{D}, \nonumber \\
  r_* &=& r+2\mathcal{M} \ln \frac{r-2\mathcal{M}}{2\mathcal{M}}.
\end{eqnarray}
Here $D$ is an arbitrary constant of dimension length and $\bar{V}_{\ell}
=(1-2\mathcal{M}/r)V_{\ell}$.

The derivation of the Zerilli function for given values of
$\ell$ and $m$ from a non-linear
numerical simulation is performed as follows. We first calculate the
physical 4-metric from the BSSN-variables and interpolate onto a
sphere of constant extraction radius $r_{\rm ex}$. Next we transform
from Cartesian to spherical metric components.
The key step is to use the
orthogonality relations of the tensor harmonics which enable us
to separate the contributions of the individual multipole moments
from the metric.
A straightforward albeit lengthy calculation shows that the perturbation
functions are related to the spherical 4-metric components by
\begin{eqnarray}
  H^{\ell m}_0 &=& \int g_{tt} Y^*_{\ell m} d\Omega, \nonumber \\
  H^{\ell m}_1 &=& \int g_{tr} Y^*_{\ell m} d\Omega, \nonumber \\
  H^{\ell m}_2 &=& \int g_{rr} Y^*_{\ell m} d\Omega, \nonumber \\
  h^{\ell m}_0 &=& \frac{1}{\ell(\ell+1)}
         \int g_{t\theta}\,\,\partial_{\theta}
         Y^*_{\ell m} + \frac{g_{t\phi}}{\sin^2\theta}\,\,\partial_{\phi}
         Y^*_{\ell m} d\Omega \nonumber \\
  h^{\ell m}_1 &=& \frac{1}{\ell(\ell+1)}\int
         g_{r\theta}\,\,\partial_{\theta}
         Y^*_{\ell m} + \frac{g_{r\phi}}{\sin^2\theta}\,\,\partial_{\phi}
         Y^*_{\ell m} d\Omega \nonumber \\
  G^{\ell m} &=& \frac{1}{r^2\,\ell(\ell+1)[\ell(\ell+1)-2]} \nonumber \\
     &&  \int \left(
         g_{\theta \theta}-\frac{g_{\phi \phi}}{\sin^2\theta}
         \right)\,W^*_{\ell m} + 2\frac{g_{\theta \phi}}{\sin\theta}
         \frac{X^*_{\ell m}}{\sin\theta} d\Omega, \nonumber \\[10pt]
  K^{\ell m} &=& \frac{\ell(\ell+1)}{2}G^{\ell m} \nonumber \\
     &&  + \frac{1}{2r^2} \int \left(
         g_{\theta\theta} + \frac{g_{\phi\phi}}{\sin^2\theta}
         \right) Y^*_{\ell m} d\Omega,
\end{eqnarray}
where a superscript $^*$ denotes the complex conjugate,
\begin{eqnarray}
  X_{\ell m} &=& 2(\partial_{\theta} - \cot \theta)\partial_{\phi}
                 Y_{\ell m} \\
  W_{\ell m} &=& \left(\partial_{\theta \theta} - \cot \theta \partial_{\theta}
                 - \frac{1}{\sin^2\theta}\partial_{\phi \phi} \right)
                 Y_{\ell m},
\end{eqnarray}
and
the integration is performed over the sphere of constant extraction
radius. In order to also calculate the derivatives in Eq.\,(\ref{eq: q1q2}),
we extract the perturbation functions at radii $r_{\rm ex}-dr$,
$r_{\rm ex}$ and $r_{\rm ex}+dr$ on every time slice which
contains the corresponding refinement level. The Zerilli function then
follows directly from Eqs.\,(\ref{eq: q1q2}), (\ref{eq: zerilli}).

We have tested the wave extraction with the analytic solution of
Teukolsky \cite{Teukolsky1982}, which describes a quadrupole
wave in linearized
general relativity. The background is the Minkowski metric in this
case which corresponds to the limit $\mathcal{M}=0$ in the relations of this
section.  Because the quadrupole $\ell =2$, $m=0$
is the dominant multipole for all simulations discussed in this
\begin{figure}
  \begin{center}
    \includegraphics[width=200pt,angle=-90]{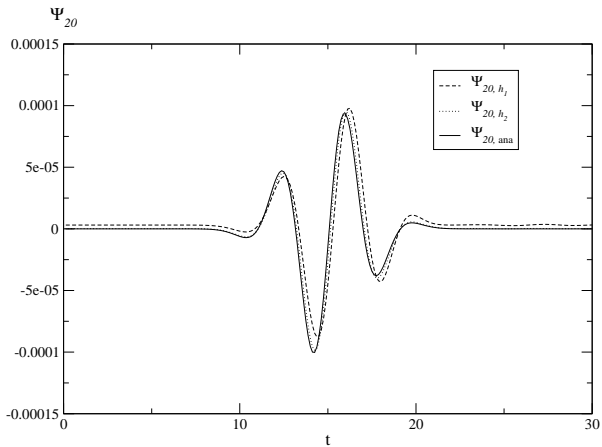}
    \caption{The numerically calculated waveforms
             (dashed and dotted curve) for the $\ell=2$,
             $m=0$ Teukolsky wave are compared with the analytic
             result (solid curve).}
    \label{fig: teuk_psi20}
  \end{center}
\end{figure}
work, we also choose the Teukolsky wave with $m=0$.
In prescribing the free function $F$
[cf.\,Eq.\,(6) in \cite{Teukolsky1982}], we follow the suggestion by
Eppley \cite{Eppley1979} and superpose an ingoing and an outgoing
wave packet. This packet is of Gaussian-like shape
\begin{equation}
  F(x) = a x e^{(-x/\sigma)^2},
\end{equation}
with $x=t\mp r$. The resulting analytic expression for the
Zerilli function $\Psi_{20}$ is rather lengthy but calculated
straightforwardly.
We evolve these data for $a=10^{-5}$ and
$\sigma=1$ in octant symmetry on a numerical domain of
size $20^3$ with two refinement levels, the fine level having half the
extent and twice the resolution of the coarse. In Fig.\,\ref{fig: teuk_psi20}
we compare the numerically computed $\Psi_{20}$ for resolutions
$h_1=0.25$ (dashed) and $h_2=0.125$
(dotted curve) in the finer refinement level.
We have also doubled the number of points on the extraction sphere from
$80\times 40$ to $160\times 80$ in the $\theta$ and $\phi$ direction.
In comparison the
solid line represents the analytic result and demonstrates the
convergence of the wave-forms. A quantitative evaluation shows that the
convergence is fourth order in the early and late stages
of this simulation and about third order in between when the wave pulse
reaches the extraction radius $r_{\rm ex}=15$. This behavior is compatible
with the fact that we use a fourth-order integration scheme for the
integration on the sphere and the main evolution code is second-order
convergent.


\ifsubmit


\else

\bibliography{Extras/uli.bib}

\fi

\end{document}